\documentclass[sigconf]{acmart}

\AtBeginDocument{%
  \providecommand\BibTeX{{%
    \normalfont B\kern-0.5em{\scshape i\kern-0.25em b}\kern-0.8em\TeX}}}

\copyrightyear{2022}
\acmYear{2022}
\setcopyright{acmcopyright}

\acmConference[SIGIR '22]{Proceedings of the 45th International ACM SIGIR Conference on Research and Development in Information Retrieval}{July 11--15, 2022}{Madrid, Spain}
\acmBooktitle{Proceedings of the 45th International ACM SIGIR Conference on Research and Development in Information Retrieval (SIGIR '22), July 11--15, 2022, Madrid, Spain}
\acmISBN{978-1-4503-8732-3/22/07}
\acmPrice{15.00}
\acmDOI{10.1145/3477495.3531910}

\usepackage{multirow}
\usepackage{bbm}
\usepackage{physics}
\usepackage{caption}
\usepackage{subcaption}
\usepackage{lipsum}
\usepackage[normalem]{ulem}
\usepackage{hyperref}
\useunder{\uline}{\ul}{}

\begin{document}

\title{Exploiting Session Information in BERT-based Session-aware Sequential Recommendation}


\author{Jinseok Seol}
    \email{jamie@europa.snu.ac.kr}
    \affiliation{%
        \institution{Seoul National University}
        \city{Seoul}
        \country{Republic of Korea}
    }
\author{Youngrok Ko}
    \email{yrko1@europa.snu.ac.kr}
    \affiliation{%
        \institution{Seoul National University}
        \city{Seoul}
        \country{Republic of Korea}
    }
\author{Sang-goo Lee}
    \email{sglee@europa.snu.ac.kr}
    \affiliation{%
        \institution{Seoul National University}
        \city{Seoul}
        \country{Republic of Korea}
    }



\begin{abstract}



    In recommendation systems, utilizing the user interaction history as sequential information has resulted in great performance improvement.
    However, in many online services, user interactions are commonly grouped by sessions that presumably share preferences, which requires a different approach from ordinary sequence representation techniques.
    To this end, sequence representation models with a hierarchical structure or various viewpoints have been developed but with a rather complex network structure.
    In this paper, we propose three methods to improve recommendation performance by exploiting session information while minimizing additional parameters in a BERT-based sequential recommendation model: using session tokens, adding session segment embeddings, and a time-aware self-attention.
    We demonstrate the feasibility of the proposed methods through experiments on widely used recommendation datasets.

\end{abstract}

\ccsdesc[500]{Information systems~Recommender systems}

\keywords{
    Session-aware Recommendation,
    Sequential Recommendation,
    Temporal Self-Attention
}

\maketitle



\section{Introduction}

    After the great success of the collaborative filtering algorithm \cite{sarwar2001item}, the recommendation system has gone through another breakthrough with leverage of sequential information \cite{hidasi2015session, quadrana2018sequence, kang2018self, sun2019bert4rec}.
    Models using RNN \cite{hidasi2015session} or attention mechanism \cite{kang2018self, sun2019bert4rec}, which has been successfully studied in natural language processing \cite{vaswani2017attention, devlin2018bert}, recommend products that fit user taste by understanding and analyzing the user's interaction history as sequence information.
    However, when using a recommendation system in real-life service, a sequence encoding technique different from natural language processing is required because user activities are performed in the unit of sessions \cite{quadrana2017personalizing}.
    As one of the research directions, session-based recommendation, which recommends using only a relatively short history of the current session, has been widely studied \cite{hidasi2015session, li2017neural, liu2018stamp, wu2019session, cho2021unsupervised}.
    Meanwhile, in the case with known user history, the session-aware model is also being studied, that is, a method of making a recommendation considering the user's long-term preference and short-term preference at the same time \cite{ma2020temporal, zhang2020personalized, xu2021long, latifi2021session}.
    In this case, the sequence is grouped through hierarchy or represented into graphs \cite{ye2020cross, wang2020next, zhang2020adaptive, fan2021continuous, sun2022sequential, zhang2022rethinking}.
    Moreover, temporal information can be induced as part of inductive bias \cite{ma2020temporal, li2020time, cho2020meantime} or various user intents can be considered \cite{zheng2019balancing, ren2020sequential, cen2020controllable, wang2020toward, wu2021rethinking, yu2021graph, zhang2022dynamic, chen2022intent,  zang2022mpan}.
    However, these session-aware methods commonly come with a rather complex model structure in order to explicitly represent the session information.

    On the other hand, in the case of the BERT model \cite{vaswani2017attention, devlin2018bert}, which is having great success in natural language processing, the model is trained with a masked language model (MLM) using a learnable mask token.
    In addition, to perform the task of inferring the relationship between two sentences \cite{conneau2017supervised, reimers2019sentence, he2020deberta}, the input sentences are distinguished through sentence segment embeddings.
    When applying BERT in the recommendation system \cite{sun2019bert4rec}, it is difficult to take advantage of the pretraining using MLM, but these components still have room for utilization in recommendation models \cite{de2021transformers4rec, zhang2021language}.

    As another element of sequential recommendation, temporal information can be used, which normally does not exist in other sequence modeling tasks \cite{xu2019self, ye2020time, zhang2021temporal}.
    In the interaction history of the user, not only the order of the item but also the time the interaction was performed is important \cite{ye2020time}. By exploiting the temporal information, it is possible to take long-term and short-term interests into account and analyze preferences more thoroughly. 
    For time information, usually expressed as unix timestamp, feature engineering is usually required due to a unit problem \cite{lei2019tissa, wang2019regularized}, is often transformed into triangular functions based on Bochner's Theorem \cite{xu2019self} so that it can be used for vector calculation.
    Recently, using learnable parameters for temporal embedding is proposed to minimize feature engineering \cite{fan2021continuous} which can utilize absolute and relative time information as needed.

    In this paper, taking ideas from the aforementioned components, we propose three methods to embed session-awareness into a model with minimum additional model structure: using session tokens, adding session segment embeddings, and a temporal self-attention.
    We conduct experiments on widely used recommendation datasets, namely MovieLens and Steam \cite{kang2018self} and demonstrate the feasibility of the proposed methods in terms of recommendation performance gain compared to additional parameters required.


\section{Related Work}

    
    \subsection{Sequential Recommendation}
    
        The sequential recommendation model has been developed by applying the models that have been successful in the NLP field.
        RNN-based \cite{hidasi2015session, li2017neural, quadrana2018sequence}, CNN-based \cite{tang2018personalized}, Attention-based \cite{kang2018self, huang2018csan, zhang2019feature, sun2019bert4rec, wu2020sse, zhangdeep, han2021casr, fan2022sequential} have been presented. Detailed model improvements have been made to suit each environment \cite{huang2018improving, liu2018stamp, ma2019hierarchical, cho2021unsupervised, qiu2021exploiting}.
    
    
    \subsection{Session-aware Recommendation}
    
        Unlike session-based, session-aware recommendation is a method that uses the relationship between sessions for each user and makes recommendations by structurally decoupling long-term and short-term preferences from a slightly more diverse perspective \cite{latifi2021session}.
        There are models that reflect the hierarchical structure \cite{quadrana2017personalizing, wu2017session, phuong2019neural, cui2019hierarchical}, the relationship between multiple sessions \cite{ren2019repeatnet, tang2019towards, zhu2020modeling, zhang2020personalized, li2021lightweight, xu2021long}.
    
    
    \subsection{Time-aware Models}
    
        Various models have been developed to utilize temporal information with continuous values such as unix timestamp in sequence data.
        There are models that construct temporal embedding \cite{xu2019self, ye2020time, zhang2021temporal, shabani2022multi} and apply to the real environment \cite{lei2019tissa, cho2020meantime, filipovic2020modeling}, and improve sequence encoding performance by reflecting time difference to attention \cite{wang2019regularized, li2020time, ma2020temporal}.
    


\begin{figure}[t]
    \centering
    \includegraphics[width=\linewidth]{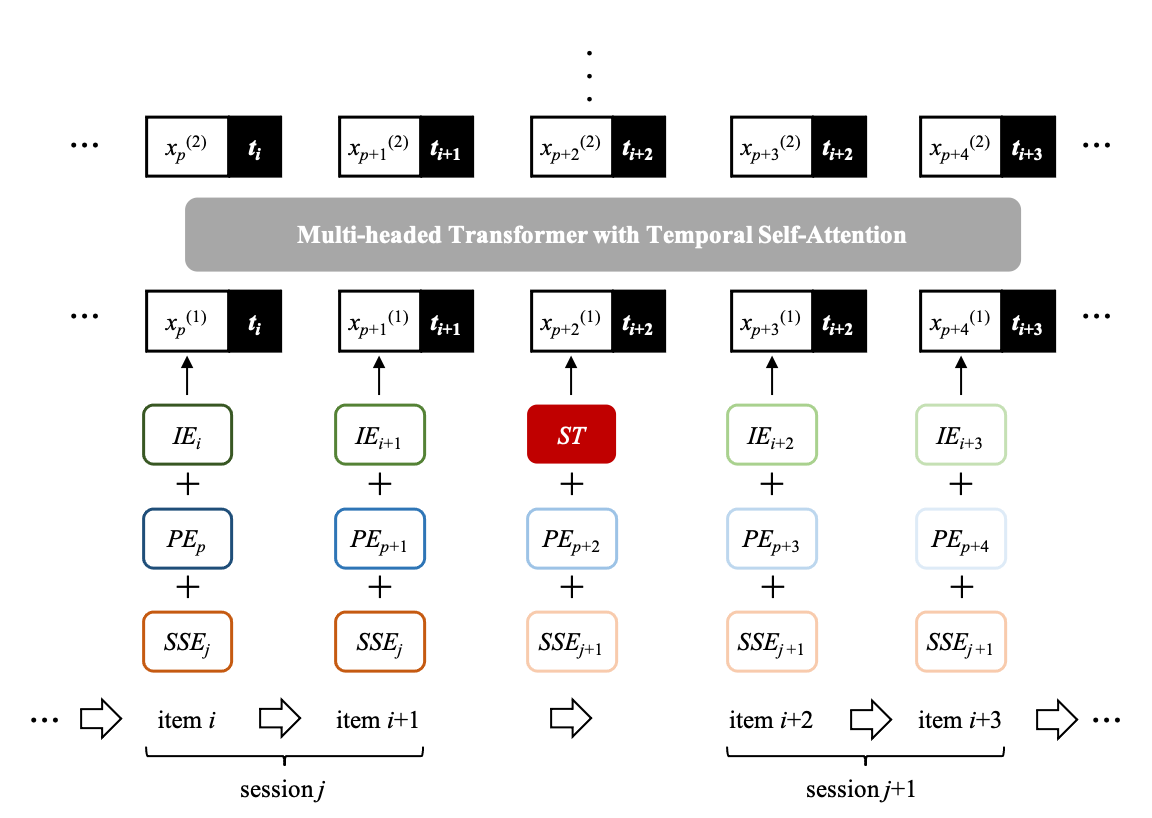}
    \caption{Input layer from model architecture showing proposed methods: session token (ST), session segment embedding (SSE), and temporal self-attention (TAS).}
    \label{fig:model}
\end{figure}

\section{Approach}


    \subsection{Background}
    
        The BERT-based sequential recommendation model is trained through the masked language model \cite{devlin2018bert}.
        First, as model input, a user's interaction history is cut to a fixed length $L$, the item embedding is set as the token embedding, and the learnable positional embedding is added according to the position of an item.
        We stack a few Transformer \cite{vaswani2017attention} layers and predict items through the softmax classifier in the last layer.
        In the training step, a randomly selected item is replaced with a mask token, and we train the model to predict the item in the corresponding position.
        At the inference step, the last-positioned item token is set to a mask token in order to predict the next item.
    
    
    \subsection{Session Token}
    
        The simplest way to distinguish sessions in the sequence of items is to insert a delimiter between item sequences \cite{ma2020temporal}.
        A learnable extra token called session token (ST) is inserted between sessions as if it is an item embedding.
        Unlike the padding token, it is not excluded from attention, and it has the effect of shifting the positional embedding by one per session.
        The advantage of this method is that it can indicate whether the input is a new session or not at inference time.
        
    
    \subsection{Session Segment Embedding}
    
        Segment embedding can be used as another way to differentiate sessions.
        When inferring the relationship between two sentences in natural language processing, segment embedding is added to the input representation to distinguish two sentences \cite{conneau2017supervised}.
        Similarly, we use learnable session segment embedding (SSE) that indicates the ordinal position of sessions, which can be thought of as another positional embedding that provides a hierarchy of sequences.
        Note that similar to the session token, this method can also indicate whether it is a new session or not at inference time.

        For $p$-th item $i$ in $j$-th session of a user, our input representation becomes: $x = IE_i + PE_p + SSE_j$ where $IE$ is an item embedding, $PE$ is a positional embedding from BERT, and $SSE$ is a session segment embedding.
        The maximum number of sessions is limited so that only the most recent $m$ sessions are considered.
        As in the implementation of positional embedding, ordinals are attached in the most recent order and padding is filled to match the model input length $L$.
    
    
    \subsection{Temporal Self-Attention}
    
        Among many temporal encoding methods for the time-aware sequential recommendation \cite{wang2019regularized, fan2021continuous, xu2019self, zhang2021temporal, cho2020meantime, ye2020time, zang2022mpan}, we adopt the learnable temporal encoding methodology from \cite{fan2021continuous} and the representation concatenation for the self-attention scheme from \cite{xu2019self}.
        With this strategy, we can consider both long or short-term preference if necessary since temporal characteristics are implicitly reflected through training the temporal parameters, rather than explicitly engineering the unix timestamp features \cite{cho2020meantime}.
        
        For a timestamp $t$, we define a temporal encoding (TE) as follows:
        \begin{equation}
            TE(t) =
                \left[
                    \cos(\omega_1 t + \theta_1)
                    \,\,
                    \cdots
                    \,\,
                    \cos(\omega_{d_T} t + \theta_{d_T})
                \right]^\top,
        \end{equation}
        where $d_T$ is a temporal dimension, and $\omega_i$, $\theta_i$ are learnable parameters.
        We concatenate temporal encoding vectors $T$ to the input representation $X$, which gives us a temporal self-attention (TAS) as follows:
        \begin{equation}
            TAS(X, T) =
                \text{softmax}
                \left(
                    \frac{
                        [X \,\, T][X \,\, T]^\top
                    }{
                        \sqrt{d_X + d_T}
                    }
                \right)
                X
            ,
        \end{equation}
        where $d_X$ is an input dimension of $X$.
        Here we can see that the attention weight $a_{ij}$ between $(x_i, t_i)$ and $(x_j, t_j)$ is calculated as
        \begin{equation}
            a_{ij} = x_i^\top x_j + TE(t_i)^\top TE(t_j),
        \end{equation}
        so that the weight becomes sum of self-attentiveness and temporal attentiveness \cite{xu2019self}.
        
        For multi-layered and multi-headed Transformer layers, we concatenate TE on each layer and head.
        Note that TE can be trained on each layer or head separately, but empirically no significant improvements were found.
    

    \subsection{Model Architecture}
        
        The input representation layer including all proposed methods is shown in Figure \ref{fig:model}.
        The rest part of the model is identical to BERT4Rec \cite{sun2019bert4rec}.
        Note that the difference from SASRec \cite{kang2018self}, which uses an autoregressive decoder, is that information other than item embedding such as positional embedding, session segment embedding, and temporal encoding can be utilized at inference time for the to-be-predicted item.
    


\begin{table}
    \caption{Dataset statistics after preprocessing. $Q_i$ denotes $i$-th quantile.}
    \label{tab:dataset}
    \begin{tabular}{cc|rrr}
        \toprule
            \multicolumn{2}{c|}{Dataset}
            &  Steam  &   ML-1M  &    ML-20M  \\
        \midrule
            \multicolumn{2}{c|}{\#users}
            &  6,330  &   1,196  &    23,404  \\
            \multicolumn{2}{c|}{\#items}
            &  4,331  &   3,327  &    12,239  \\
            \multicolumn{2}{c|}{\#rows}
            & 49,163  & 158,498  & 1,981,866  \\
            \multicolumn{2}{c|}{density}
            &  0.18\% &   3.98\% &     0.69\% \\
        \midrule
            item/user    & $Q_1$/$Q_2$/$Q_3$ & 5/6/8 & 73/137/200 & 35/68/124 \\
            session/user & $Q_1$/$Q_2$/$Q_3$ & 2/2/2 &      2/2/3 &     2/2/3 \\
            item/session & $Q_1$/$Q_2$/$Q_3$ & 2/3/3 &    8/26/70 &   6/15/39 \\
        \bottomrule
    \end{tabular}
\end{table}

\section{Experiment}


    \subsection{Experimental Design}
    

        \subsubsection{Datasets and preprocessing}
            
            We evaluate our approach on three real-world recommendation datasets: MovieLens 1M and 20M and Steam \cite{kang2018self}.
            In the preprocessing step, similar to \cite{kang2018self}, each review or rating information was considered as an implicit positive interaction.
            For the quality of the dataset, items with an interaction count of 5 or less were removed, and users with a history length of 5 or less were also removed.
            For time-related information, unix timestamp units are used, and as in \cite{cho2021unsupervised}, sessions are separated if there is no activity for 1-day.
            In order to create a multi-session environment, the number of sessions per user is limited to 2 or more, the number of items configured per session is 2 or more, and only 200 recent items are used.
            The statistics of datasets after preprocessing are shown in Table \ref{tab:dataset}.
            For each user, the last item was used as a test item, the second most recent item as validation, and the rest as a training set.
            Note that items in the validation were also enclosed when testing.
        
        
        \subsubsection{Evaluation metrics}
        
            As evaluation metrics, Recall and NDCG were used \cite{he2017neural}.
            For each user, 100 negative items not in interaction history were randomly sampled and ranked together with the ground-truth item.
            As is known from \cite{krichene2020sampled}, random negatives are biased samplers, so we additionally construct \emph{popular negatives} and \emph{all negatives}.
            Popular negatives take sorted items based on popularity as negative items, which act as a kind of ``hard negatives''.
            In the case of all negatives, the entire item including the user's positive items is used as a candidate set and ranked together.
            In a real-world commercial recommendation system, post-processing is often conducted after selecting recommendation candidate items, thus all negatives metric presumably imitates real-world performance.

        
        \subsubsection{Implementation details}
        
            We use PyTorch implementation of the standard BERT implementation \cite{sun2019bert4rec} and trained from scratch using the AdamW \cite{loshchilov2018fixing} optimizer on a Titan Xp GPU.
            All hyper-parameters were tuned through grid search, and we report the one with the best performance in the final result.
            Hidden dimension is searched among [8, 16, 32, 64, 128, 256], mask probability from [0.05, 0.1, 0.15, 0.2, 0.25, 0.3], number of attention head from [1, 2, 4, 8], max session segments from [2, 3, 4, 5], and temporal dimension from [8, 16, 32, 64].
            Based on dataset statistics, the max sequence length is set to 200, 100, and 15 for ML-1M, ML-20M, and Steam respectively.
            
            All source codes including model implementation, dataset preprocessing, experiments with hyper-parameter tuning is released to the public in \url{https://github.com/theeluwin/session-aware-bert4rec}.
        
    

\begin{figure}[t]
    \centering
    \begin{subfigure}[b]{0.5\linewidth}
        \centering
        \includegraphics[width=\textwidth]{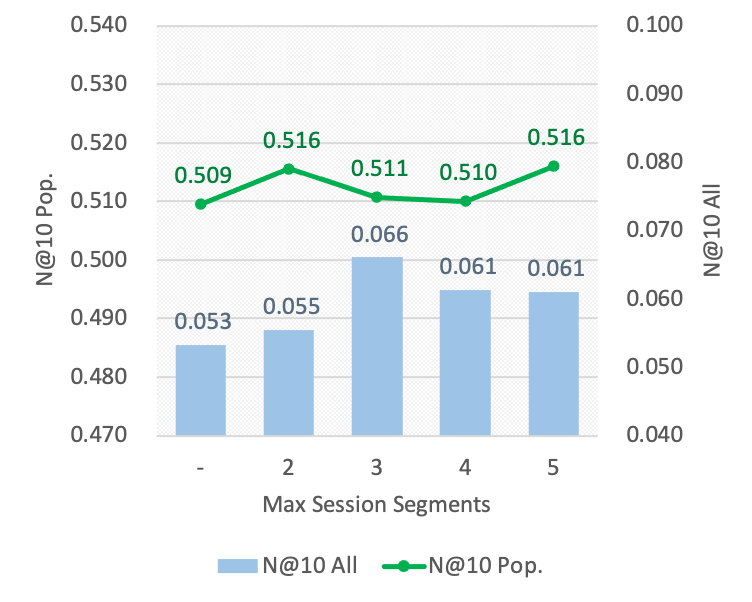}
        \caption{Max Session Segments}
        \label{fig:HP-SSE}
    \end{subfigure}%
    \begin{subfigure}[b]{0.5\linewidth}
        \centering
        \includegraphics[width=\textwidth]{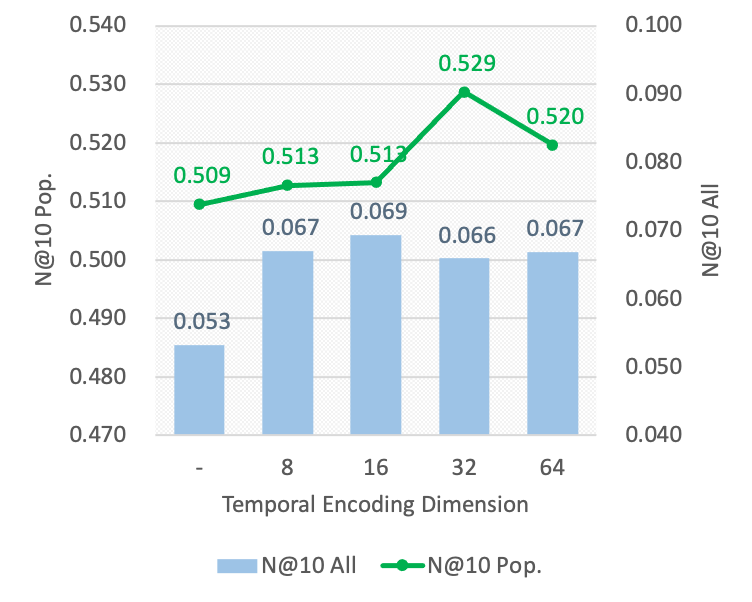}
        \caption{Temporal Dimension}
        \label{fig:HP-TSA}
    \end{subfigure}
    \caption{Impact of hyper-parameters on ML-1M dataset.}
    \label{fig:HP}
\end{figure}

\begin{table*}[t]
    \caption{Recommendation performance comparison with ablation study on various evaluation metrics. Bold and underline text indicate the best and the second-best score, respectively.}
    \label{tab:ablation}
    \begin{tabular}{c|l|r|rr|rr|rr}
        \toprule
            Dataset
                    & Model            & \#params  & R@10 Ran.       & N@10 Ran.       & R@10 Pop.       & N@10 Pop.       & R@10 All        & N@10 All        \\
        \midrule
            \multirow{10}{*}{Steam}
                    & SASRec           & 5,385,708 & 0.7834          & 0.6726          & 0.5523          & 0.5007          & 0.5164          & 0.4610          \\
                    & BERT4Rec         & 5,385,708 & 0.7987          & 0.6915          & 0.5670          & 0.5196          & 0.5313          & 0.4782          \\
                    & + ST             & 5,385,964 & 0.8063          & 0.7080          & \textbf{0.6033} & \textbf{0.5602} & 0.5578          & \textbf{0.5199} \\
                    & + SSE            & 5,386,988 & 0.8060          & 0.7045          & 0.5919          & 0.5482          & 0.5528          & 0.5138          \\
                    & + TSA            & 5,389,996 & 0.8079          & \textbf{0.7129} & 0.5997          & 0.5562          & 0.5577          & 0.5179          \\
                    & + ST + SSE       & 5,387,244 & 0.8011          & 0.7031          & 0.5978          & 0.5543          & 0.5558          & 0.5147          \\
                    & + ST + TSA       & 5,390,252 & {\ul 0.8120}    & 0.7093          & {\ul 0.6030}    & {\ul 0.5596}    & \textbf{0.5616} & {\ul 0.5187}    \\
                    & + SSE + TSA      & 5,391,276 & \textbf{0.8125} & {\ul 0.7103}    & 0.5997          & 0.5551          & 0.5566          & 0.5181          \\
                    & + ST + SSE + TSA & 5,391,532 & 0.8093          & 0.7078          & 0.5979          & 0.5486          & {\ul 0.5581}    & 0.5118          \\
        \midrule
            \multirow{10}{*}{ML-1M}
                    & SASRec           & 4,918,016 & 0.7199          & 0.4962          & 0.4189          & 0.2674          & 0.1480          & {\ul 0.0742}    \\
                    & BERT4Rec         & 4,918,016 & 0.7341          & 0.5100          & 0.5025          & 0.3211          & 0.1129          & 0.0508          \\
                    & + ST             & 4,918,272 & {\ul 0.7508}    & 0.5160          & 0.4900          & 0.3271          & 0.1338          & 0.0618          \\
                    & + SSE            & 4,918,784 & 0.7383          & 0.5260          & 0.5109          & {\ul 0.3348}    & 0.1455          & 0.0678          \\
                    & + TSA            & 4,919,136 & 0.7408          & 0.5259          & 0.4933          & 0.3239          & {\ul 0.1522}    & \textbf{0.0790} \\
                    & + ST + SSE       & 4,919,040 & \textbf{0.7533} & 0.5230          & {\ul 0.5134}    & \textbf{0.3362} & 0.1497          & 0.0739          \\
                    & + ST + TSA       & 4,919,392 & 0.7441          & 0.5256          & 0.5125          & 0.3328          & \textbf{0.1564} & 0.0728          \\
                    & + SSE + TSA      & 4,919,904 & 0.7492          & {\ul 0.5270}    & 0.4900          & 0.3086          & 0.1513          & 0.0704          \\
                    & + ST + SSE + TSA & 4,920,160 & 0.7458          & \textbf{0.5298} & \textbf{0.5217} & 0.3283          & 0.1505          & 0.0701          \\
        \midrule
            \multirow{10}{*}{ML-20M}
                    & SASRec           & 9,464,272 & 0.9014          & 0.6954          & 0.4370          & 0.2839          & 0.1489          & \textbf{0.0807} \\
                    & BERT4Rec         & 9,464,272 & 0.9053          & 0.6944          & 0.4729          & 0.3051          & 0.1381          & 0.0724          \\
                    & + ST             & 9,464,528 & 0.9058          & 0.6958          & \textbf{0.4985} & \textbf{0.3287} & 0.1436          & 0.0750          \\
                    & + SSE            & 9,465,552 & 0.9070          & 0.6975          & 0.4711          & 0.3079          & 0.1450          & 0.0740          \\
                    & + TSA            & 9,481,040 & {\ul 0.9114}    & 0.7010          & 0.4690          & 0.3025          & 0.1439          & 0.0754          \\
                    & + ST + SSE       & 9,465,808 & 0.9069          & 0.6967          & 0.4845          & 0.3183          & 0.1418          & 0.0735          \\
                    & + ST + TSA       & 9,481,296 & 0.9105          & 0.7024          & {\ul 0.4960}    & {\ul 0.3258}    & \textbf{0.1499} & {\ul 0.0790}    \\
                    & + SSE + TSA      & 9,482,320 & 0.9111          & {\ul 0.7029}    & 0.4847          & 0.3202          & 0.1447          & 0.0756          \\
                    & + ST + SSE + TSA & 9,482,576 & \textbf{0.9114} & \textbf{0.7029} & 0.4921          & 0.3211          & {\ul 0.1492}    & 0.0786          \\
        \bottomrule
    \end{tabular}
\end{table*}

    \subsection{Recommendation Performance}
    
        As shown in Table \ref{tab:ablation}, all of our proposed methods outperform the baseline.
        Although combinations of our methods do not necessarily help the task performance, each method encourages at least some aspects of the evaluation metric or dataset.
        With proper selection of session information exploitation, we can get significant performance gain with a very little increment of the number of model parameters.
        More specifically, in the case of a session token, the parameters are only added as much as one additional item.
        In the case of session segment embedding, the parameters are increased by the number of segments, which are not many in most cases.
        For the temporal embedding, only twice as many additional parameters as the temporal dimension are used.
        Altogether, only about 0.1\% of additional parameters are used, which is  minor compared to the parameters that make up the transformer layers and item embeddings.
        Meanwhile, performance on NDCG with all negatives is improved from 8\% to 50\% or more in BERT4Rec.
    

    \subsection{Impact of Hyper-parameters}
    
        We also conducted experiments on the impact of session-related hyper-parameters, namely max session segments and temporal dimension, as shown in Figure \ref{fig:HP}.
        The performance between popular negatives and all negatives was expected to have a high correlation, but the experimental results show that the optimal hyper-parameter is slightly different depending on the evaluation metric.
        In the case of max session segments (Figure \ref{fig:HP-SSE}), a significant performance gain in all negatives is achieved only when the value exceeds a certain level, as expected in Table \ref{tab:dataset}.
        For the case of the temporal dimension (Figure \ref{fig:HP-TSA}),  hyper-parameter tuning for each dataset is required, which can be seen as a drawback as mentioned in the experimental results of \cite{fan2021continuous}.
    


\section{Limitations and Future Works}

    In this paper, we demonstrated exploiting session information for BERT-based, session-aware sequential recommendations through a learnable session token, session segment embedding, and temporal self-attention.
    Although experiment results show the potential of the proposed methods, there are limitations on datasets with different statistical characteristics related to the number of sessions per user and the number of items per session.
    Specifically, additional experiments were performed on datasets including Amazon reviews, LastFM, RetailRocket, and Diginetica, but it was difficult to acquire performance gain, which is presumed to be due to differences in data sparsity and session-related statistical characteristics.
    As an ongoing work, we will proceed and develop novel methodologies to utilize session information more thoroughly and consistently in session-aware sequential recommendation tasks.


\section*{Acknowledgments}

    This work was made in collaboration with Seoul National University and IntelliSys Co., Ltd.
    Also, this work was partly supported by Institute of Information \& communications Technology Planning \& Evaluation (IITP) grant funded by the Korean government (MSIT) [No.2021-0-00302, AI Fashion Designer: Mega-Trend and Merchandizing Knowledge Aware AI Fashion Designer Solution], [No.2021-0-02068, Artificial Intelligence Innovation Hub (Artificial Intelligence Institute, Seoul National University)], [No.2021-0-01343, Artificial Intelligence Graduate School Program (Seoul National University)].


\bibliographystyle{ACM-Reference-Format}
\bibliography{main}


\end{document}